%% file: hnlmu.tex
\documentclass[a4paper,11pt]{article}
\pdfoutput=1 

\usepackage{jheppub} 

\usepackage{slashed}
\usepackage{graphicx}
\usepackage{adjustbox}
\usepackage{subfig}
\usepackage{xcolor}
\usepackage{amsmath}
\usepackage{booktabs}

\usepackage{lineno}

\usepackage[T1]{fontenc} 
\newcommand{\mhn}{\ensuremath{M_{N_1}}}
\newcommand{\rvxp}{\ensuremath{D_{xy}^{vxp}}}
\newcommand{\mgev}{\ensuremath{\,\textrm{GeV}} }
\newcommand{\egev}{\ensuremath{\,\textrm{GeV}}}

\newcommand{\PGm}{\ensuremath{\mu N}}

\title{\boldmath Searches for Heavy Neutral Leptons at FCC-ee in final states including a muon.}

\author[a]{L Bellagamba,}
\author[b,1]{G Polesello,}
\author[b]{N. Valle.} \note{Corresponding author.}

\affiliation[a]{INFN Sezione di Bologna, Viale C. Berti Pichat 6/2, 40127 Bologna, Italy}
\affiliation[b]{INFN Sezione di Pavia, Via Bassi 6, 27100 Pavia, Italy}

\emailAdd{lorenzo.bellagamba@cern.ch}
\emailAdd{giacomo.polesello@cern.ch}
\emailAdd{nicolo.valle@cern.ch}

\abstract{The sensitivity of the CERN FCC-ee collider to the production of  heavy neutral leptons (HNL) is investigated. The study focuses on a simplified model with a single low-mass HNL mixing with a muon, and addresses the fully leptonic and semileptonic decay modes of the HNL. Complete Monte Carlo analyses of signal and background based on a parametrised detector simulation  are performed for the FCC-ee run at the $Z$-pole, resulting in an estimate of the intervals of the mixing parameter for which the FCC-ee will have a 95\% CL sensitivity  as a function of the HNL mass. 
}

\begin{document} 
\maketitle
\flushbottom

\section{Introduction}
\label{sec:intro}
\input{intro}

\section{The simulation setup}\label{sec:simulation}
\input{simulation}

\section{Fully leptonic muon channel with $\mu^+\mu^-\nu\bar{\nu}$ final state}
\input{mumu_simu}
\input{mumu_analysis}

\section{Semi-leptonic muon channel with $\mu\nu jj$ final state}
\input{mujj_reco.tex}
\input{mujj_ana.tex}

\section{Combined results and conclusions}
\input{combined.tex}
\clearpage

\acknowledgments
The work of N.V. has received funding from the European Union’s Horizon 2020 Research and Innovation programme under  GA no 101004761
\bibliography{hnlmu}
\end{document}

%% file: intro.tex
The next generation of proposed $e^+e^-$ circular colliders, such
as the CERN FCC-ee  \cite{FCC:2018evy}, will provide access to 
a broad range of physics studies.

The production of HNLs has been identified as one of the most promising new physics channels for FCC-ee at the Z pole in a seminal paper of 2014 \cite{Blondel:2014bra}. Several different decay channels and lifetime scenarios will be 
accessible at the FCC, yielding severe requirements on the performance of the detectors, which must be quantified through detailed studies taking into account realistic performance figures for the foreseen detectors. Detailed 
discussions of previous experimental studies targeting FCC-ee
are contained in \cite{Blondel:2022qqo, Abdullahi:2022jlv}.

The present study aims to evaluate the sensitivity for the production of HNLs to the $e^+e^-$ future collider, at the centre of mass energy $\sqrt{s}=91.2\egev$ and with  target integrated luminosity $L_{int}= 2.0\times 10^8\, \mathrm{pb}^{-1}$, distributed over three energy points around the $Z$ peak corresponding to $6\times10^{12}$ $e^+e^-\to Z$ interactions. A HNL mass region ranging from $5 \egev$ to $85 \egev$ is investigated. 

\begin{figure}[tbp]
\centering 
\includegraphics[width=.95\textwidth]{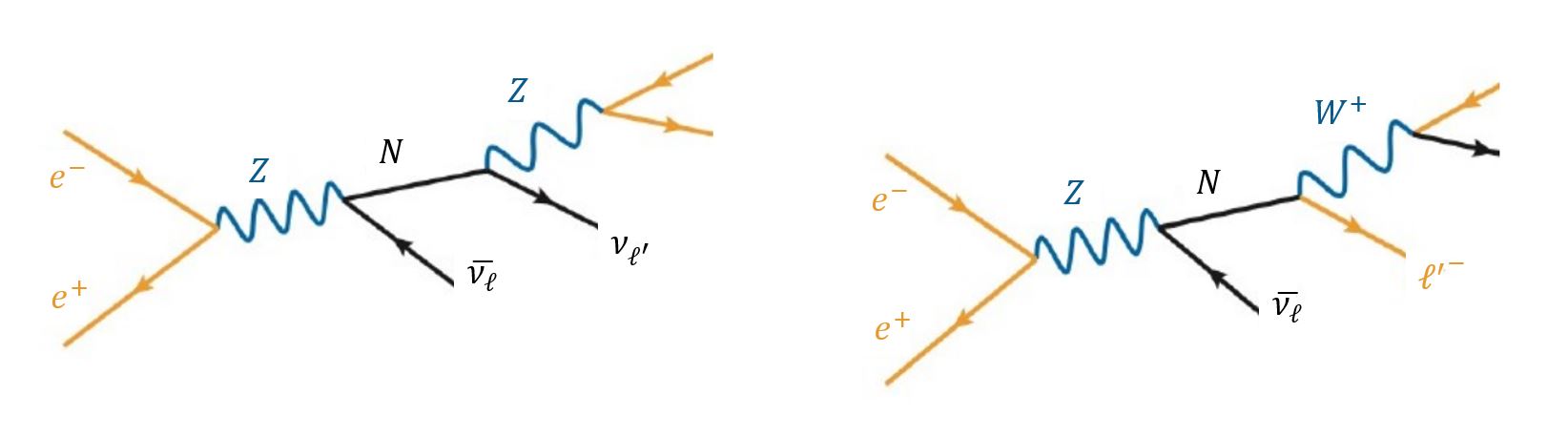}
\caption{\label{fig:HNL} Diagrams for production and decay of a heavy neutral
lepton in the decay of a $Z$ boson.}
\end{figure}

We study the production of HNL in $Z$ decay through mixing with light neutrinos.
The HNLs decay to a virtual or real $Z$ or $W$ vector boson and, respectively a light neutrino or a lepton. The vector boson decays in turn to two fermions, as shown in the diagrams of Figure~\ref{fig:HNL}.
We concentrate on a benchmark model with a single light HNL mixing only with muons, so the model is defined in terms of two parameters: the HNL mass ($\mhn$) and the mixing parameter $U_{\PGm}$ with the active neutrino. Prompt and long-lived signatures are both possible, depending on the value of those parameters.

Two decay channels are considered:
\begin{itemize}
\item
a fully leptonic final state $\mu^+\mu^-\nu$, which yields a very clean final state
signature, and is produced both through the mediation of a charged and a neutral 
vector boson. The branching fraction is approximately 5\% over the parameter range
of interest.
\item
The semileptonic decay into $\mu j j'$, where $j,j'$ are jets from $q\bar{q}'$ pairs coming from the charged vector boson coupling with HNL and the muon.  This channel is the most copiously produced, with a total branching fraction of approximately 50\% over the full HNL mass range of interest. Moreover, this channel, with a single neutrino in the final state, allows the complete reconstruction both of the mass of the HNL and of the energy of the recoiling neutrino from the $Z$ decay,  providing a very strong handle for the reduction of non-resonant background sources.
\end{itemize}

The present study complements and completes the results shown in the Snowmass reports \cite{Blondel:2022qqo, Abdullahi:2022jlv}, which focused on long-lived signatures, and for which the results including consideration of experimental effects were limited to leptonic decays of the HNL.  The experimental sensitivity of a future circular $e^+e^-$ collider to the  prompt HNL decay in $\mu j j'$  has been studied in a master thesis~\cite{nielsenth} and in two papers focused on the CEPC collider in China~\cite{Ding:2019tqq, Shen:2022ffi}. We improve on the previous work by performing a full  Monte Carlo study based on a parametrised simulation of the expected performance for the IDEA detector proposal \cite{IDEAStudyGroup:2025gbt}. 
Background studies rely on the centrally produced large background statistics for the FCC Physics-Experiment-Detector (PED)  studies \cite{winter2023samples}. Both prompt and long-lived (LLP, long-lived particles) signatures are studied in an integrated environment, which offers the possibility of statistically combining the different channels studied.

The paper is organised as follows. We first describe the generation of signal and background followed by a description of detector simulation and event reconstruction, with special focus on the reconstruction
of vertices in the inner detector and of hadronic jets. 
In the following two sections, the fully leptonic and semi-leptonic analyses are described separately, arriving at the definition of optimal signal regions.
Based on these selections, the covered area for each analysis in the model parameter space is assessed. Finally, the results of the different analyses are combined and compared to existing experimental limits.

%% file: simulation.tex

The discovery potential for Heavy Neutral Leptons with the IDEA detector \cite{IDEAStudyGroup:2025gbt} at FCC-ee is studied through a detailed Monte Carlo analysis. The generated events are passed through a parametrised simulation of the detector response, and physics objects for analysis are reconstructed from the output of the simulation.
In the following, we provide some details on the event simulation procedure.

\subsection{Model definition and signal generation}
The model of interest is implemented in the 
\verb+SM_HeavyN_LO+ \cite{Atre:2009rg,Alva:2014gxa,Degrande:2016aje} package, 
and signal samples were generated with \verb+MG5aMC@NLO+ \cite{Alwall:2014hca}.
The masses of the heavy neutrinos $N_{2}$ and $N_{3}$ were set to 10 TeV, and all mixing terms were set to zero, except for the mixining $U_{\PGm}$ between the muon and the HNL ($N_{1}$) in the model. The associated production of a muon (anti)neutrino and the heavy neutrino $N_{1}$ was simulated at a centre-of-mass energy of $91.2\egev$,
with the $N_{1}$ directly decayed into the final state of interest
in the \verb+MG5aMC@NLO+ process card, so as to have the correct decay kinematics.

For the analysis addressing the semileptonic decay, a scan was performed on the mass of $N_{1}$  ($\mhn$) between 5 and 85 GeV.
For each mass a scan on $U_{\PGm}$ was performed in a range  going from the minimal coupling yielding at least one event  decaying within 2.5 meters of the centre of the detector for the full FCC-ee Z-pole statistics, to $U^2_{\PGm}=5\times10^{-4}$ which is excluded by existing experiments. A total of 10k events were generated for each sample.

For the fully leptonic case, a scan was performed over $\mhn$ between 10 and 50 GeV with a step of 5 GeV and for couplings  $U_{\PGm}$ in the range from $10^{-2}$ to $5\times 10^{-6}$. For each sample 2k events were generated.

The LHE files generated with \verb+MG5aMC@NLO+ were hadronised with \verb+PYTHIA8+ \cite{Sjostrand:2014zea} and then fed into  the \verb+DELPHES+ \cite{deFavereau:2013fsa} fast simulation of the IDEA detector, based on the official data cards used for the  "Winter2023" production of backgrounds \cite{winter2023setup}.

For the backgrounds from $Z$ decays, the official samples produced by the central software group for the FCC PED studies under the tag ``Winter2023'' were used \cite{winter2023samples}. The total available  Monte Carlo statistics correspond to the production of approximately $3\times10^{9}$ $e^+e^-\rightarrow Z$ events.

The irreducible background from the four-fermion process
$$
e^+e^-\rightarrow \mu \nu j j
$$
was produced at LO with \verb+MG5aMC@NLO+, including both the
associated  production of a real and a virtual $W$, and $Z/\gamma$
production followed by the radiation of a virtual $W$ off one of the decay
legs of the $Z$. The only generation-level  requirements were
that leptons and jets were produced within a pseudorapidity of $\pm5$
and that the invariant mass of the two jets was in excess of 5~GeV.
The cross-section for the process is 3.2~fb, and a sample of 500k events
was produced, corresponding to approximately the full expected statistics for
the FCC-ee run at the $Z$ pole. The events were then processed through
the same \verb+PYTHIA8+-\verb+DELPHES+ chain as the signal events.

\subsection{The IDEA detector and its simulation}

The IDEA detector concept is a proposal for a general-purpose detector for the FCC-ee.
The design  includes an inner detector composed of 5 Monolithic silicon pixel (MAPS) layers
followed by a high-transparency and  high-resolution drift chamber.  
Outside the inner detector is a dual-readout electromagnetic (EM) crystal calorimeter
providing high-precision energy measurement of photons and electrons, followed
by a superconducting solenoid producing a 2~T magnetic field.
Outside the solenoid is located a dual-readout fibre calorimeter, providing, together
with the EM calorimeter, high precision energy measurement of hadrons.
The detector is completed by three $\mu$-rwell  layers for muon detection, embedded in the 
return yoke of the solenoid.

The present analysis relies on the parametrised simulation in \verb+DELPHES+ of the 
inner detector for the estimate of the tracking and vertexing performance
of IDEA, complemented by a parametrised simulation of the calorimeter response
to reconstruct hadronic jets.
%
%

The \verb+DELPHES+ simulation software relies on a full description of the geometry of the IDEA vertex detector and drift chamber and accounts for the finite detector resolution and for the multiple scattering in each tracker layer. It turns charged particles emitted within the angular acceptance of the tracker into five-parameter tracks (the helix parameters that describe the trajectory of the particle, including the transverse and longitudinal impact parameters), and determines the full covariance matrix of these parameters.
Vertices are reconstructed using these tracks as 
input, based on a simple $\chi^2$ minimisation with constraints, producing 3D
vertices with their $\chi^2$ and covariance matrix. More details on the vertexing code used here can be found in ~\cite{Bedeschi:2024uaf}.

The calorimeter response is simulated by smearing the energy of electrons and photons
based on the expected resolution of the crystal EM calorimeter, parametrised as 
$$
\frac{\sigma(E)}{E}=\frac{0.03}{\sqrt{E}}\,\oplus\,0.005\,\oplus\,\frac{0.002}{E}.
$$
The response for hadrons is parametrised as 
$$
\frac{\sigma(E)}{E}=\frac{0.3}{\sqrt{E}}\,\oplus\,0.01\,\oplus\,\frac{0.05}{E},
$$
which is the expected response for the dual readout fiber calorimeter.

From the tracks and energy depositions in the calorimeter, particle flow objects (PFOs) are built. Reconstructed tracks are used for charged particles. For neutrals, the
PFOs are vectors with magnitude equal to the calorimetric energy measurement, and direction corresponding to the segment connecting the centre of the detector with the centre of the face of the hit calorimeter cell, smeared by the size of the cell.
Hadronic jets are reconstructed giving collections of particle flow objects in input 
to the \verb+FASTJET+ package~\cite{Cacciari:2011ma}.

%% file: mumu_simu.tex
The decay of the HNL into $\mu \mu \nu_{\mu}$ can take place through the virtual
exchange of an off-shell $Z$ or a $W$ boson, and it has a branching fraction 
of $\sim5\%$ in the considered model. 
It is an experimentally clean channel with only two reconstructed muons in the detector and missing energy due to escaping neutrinos. The long-lived signature, which characterizes HNL decays in a significant portion of the parameter space, can be exploited to suppress the massive Standard Model background due to the decay of $Z$ to muons, taus and heavy flavours and to the 
four fermion process  $e^+e^-\to \mu \mu \nu \nu$.

\subsection{Event processing and reconstruction}\label{sec:mumureco}
The signal and background events are generated and reconstructed using the \verb+DELPHES+ fast simulation as described in Section~\ref{sec:simulation}. A pre-selection is applied that requires two final-state muons with momentum larger than 3 $\egev$ and nothing else in the detector. In addition, the two muon tracks are required to be fitted to a common vertex, and a cut on the $\chi^2$ per degree of freedom, $\chi^2/\mathrm{ndf} < 10$, is applied to ensure a reasonable fit quality. The distance between the reconstructed vertex and the interaction point (IP) is the crucial parameter used to separate signal from background. 

\begin{figure}[h!]%
\centering
\includegraphics[width=0.49\textwidth]{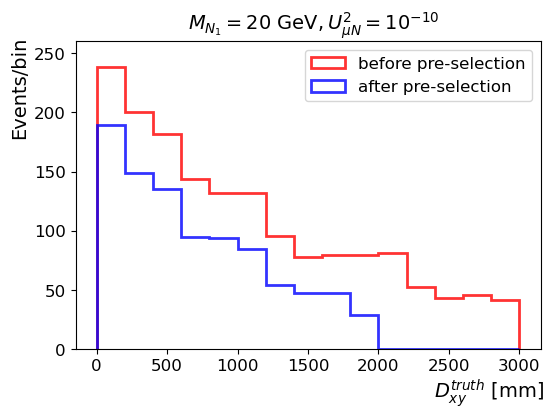}
\includegraphics[width=0.50\textwidth]{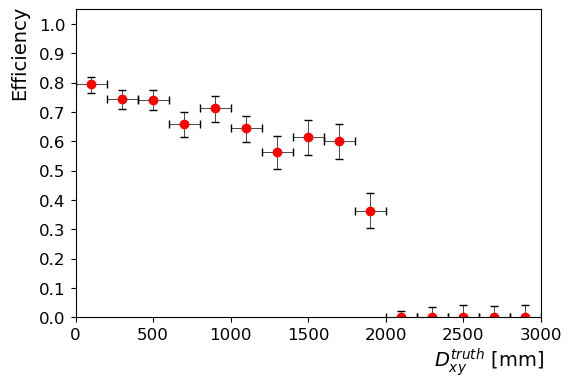}
\caption{Distribution of the transverse distance between the IP and the decay vertex at the generator level before and after the pre-selection requirements for 2k signal events generated at $\mhn=20 \egev$ and $U^2_{\PGm}=10^{-10}$ (left) and corresponding efficiency (right).}
\label{fig-eff_sig20}
\end{figure}

\begin{figure}[h!]%
\centering
\includegraphics[width=0.49\textwidth]{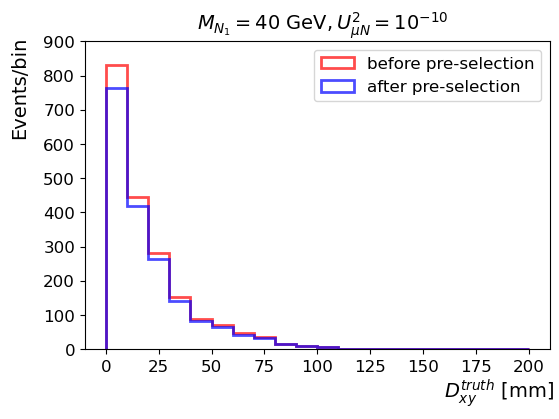}
\includegraphics[width=0.49\textwidth]{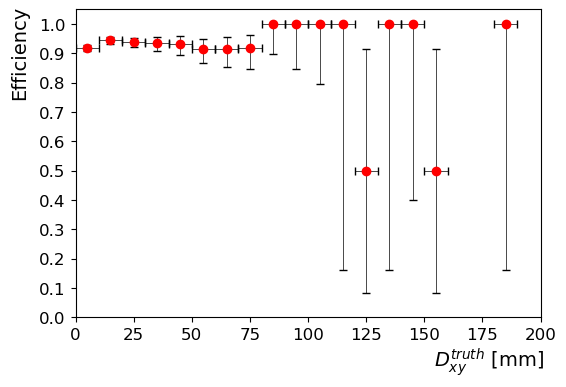}
\caption{Distribution of the transverse distance between the IP and the decay vertex at the generator level before and after the pre-selection requirements for 2k signal events generated at $\mhn=40 \egev$ and $U^2_{\PGm}=10^{-10}$ (left) and corresponding efficiency (right).}
\label{fig-eff_sig40}
\end{figure}

Figures~\ref{fig-eff_sig20} and~\ref{fig-eff_sig40} show, for two generated points, the distribution of the transverse distance between the IP and the decay vertex at generator level ($D_{xy}^{\mathrm{truth}}$), before and after the pre-selection requirements (left), together with the corresponding pre-selection efficiency (right).
At a large distance from the interaction point, the lower mass point, characterized by a much longer decay-length, exhibits a clear drop in the efficiency due to the acceptance of the central tracking detector. Muons without a track in the central tracking detector are not considered in the present analysis.

The cross-sections of the simulated background samples are several orders of magnitude larger than the signal in a large part of the parameter space, and the generated statistics is not sufficient to populate the tails of the distributions which contribute to the possible background. The samples can hence only suggest reasonable selection requirements for signal to background separation taking into account the intrinsic limitations of the study.

\begin{figure}[h!]%
\centering
\includegraphics[width=0.6\textwidth]{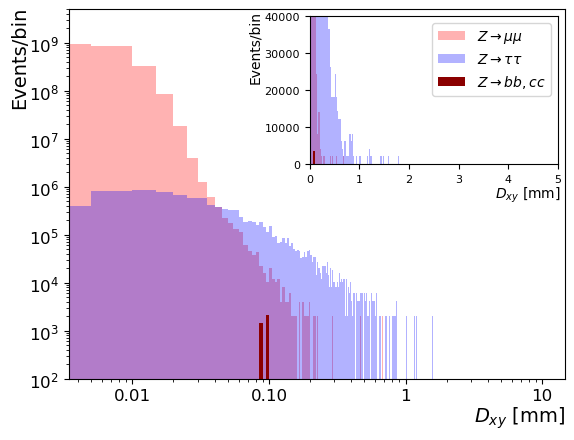}
\caption{$D_{xy}$ distribution for $Z\to \mu \mu$, $Z\to \tau \tau$ and  $Z\to bb/cc$, applying pre-selection requirements and a cut on the angle between the two reconstructed muons to reject back-to-back topology as described in the text. The inset shows the same distribution in linear scale. The plot has been normalized to a luminosity corresponding to $6 \times 10^{12}$ $Z$ events.}
\label{fig-zbkg}
\end{figure}

The natural variable to consider for background rejection, taking advantage of the LLP topology of the signal, is the reconstructed transverse distance $D_{xy}$ between the displaced vertex and the interaction point. Figure~\ref{fig-zbkg} shows the distributions of $D_{xy}$ for $Z~\to~\mu \mu/\tau \tau$ and $Z \to bb / cc$, applying the pre-selection requirements and a further cuts $\cos(\alpha_{\mu\mu})~>~-~0.95$, where $\alpha_{\mu\mu}$ is the angle between the two reconstructed muon tracks at the decay vertex. 

The requirement on $\cos(\alpha_{\mu\mu})$ has a negligible impact on the signal efficiency and removes back-to-back topologies that could produce relatively large $D_{xy}$ values due to poor vertex reconstruction.
While $Z$ decays to muons produce prompt muons, $Z$ decays to taus or heavy quarks give rise to genuinely displaced vertices, arising from leptonic tau decays, semi-leptonic heavy-quark decays, and vector meson decays.
However, as clearly shown in Figure~\ref{fig-zbkg}, the background from heavy-quark production is strongly suppressed by the exclusive requirement of two final-state muons, while the $\tau$-induced background exhibits a tail extending up to a few millimeters.

%% file: mumu_analysis.tex
\subsection{Final selection and signal sensitivity}
The strategy adopted in this study involves the variable $D_{xy}$ to separate signal from background, following a conservative approach and assuming negligible background after the final selection requirements listed below:
\begin{itemize}
    \item 2 tracks reconstructed as muons with momentum $> 3 \egev$ and nothing else in the detector;
    \item $\cos(\alpha_{\mu\mu}) > -0.95$;
    \item a reconstructed vertex with $D_{xy} > 10$ mm.
\end{itemize}
An improved background simulation will allow for future optimization of the selection, leveraging additional kinematic variables and potentially relaxing the $D_{xy}$ requirement to enhance the sensitivity in the parameter space, especially towards higher masses.

\begin{figure}[h!]%
\centering
\includegraphics[width=0.8\textwidth]{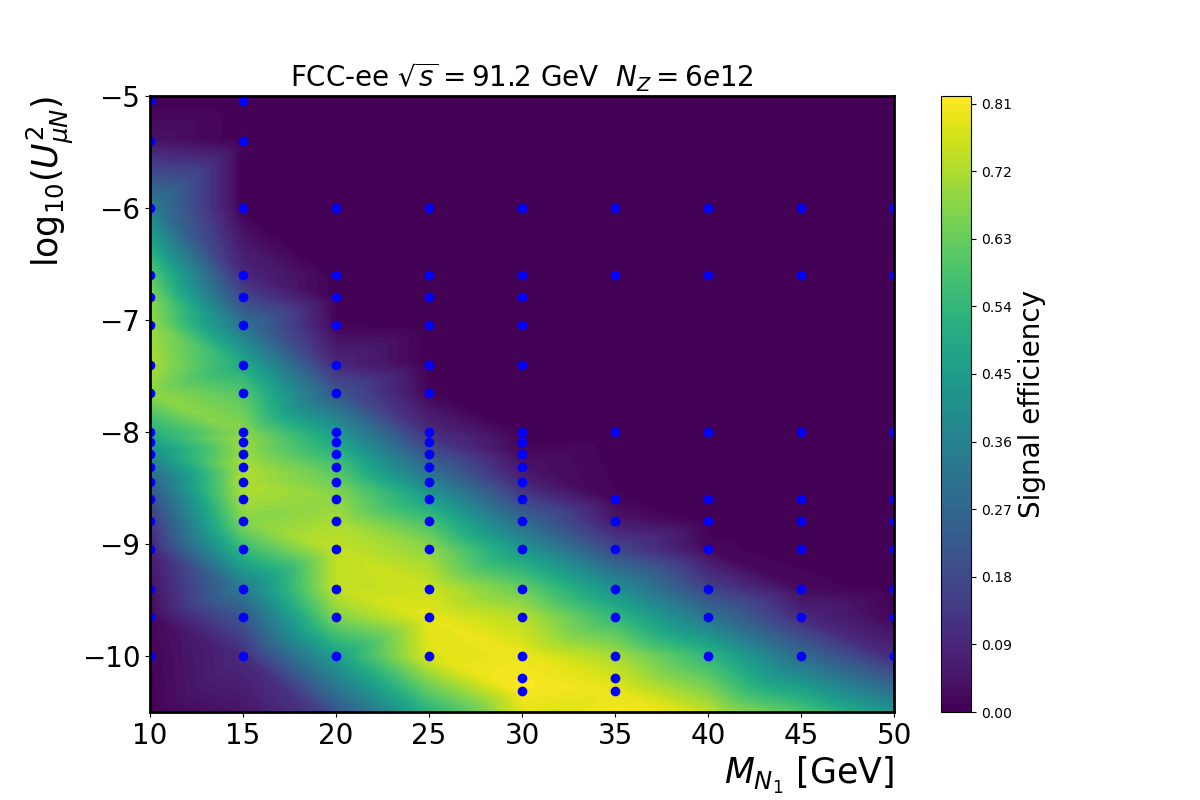}
\caption{Signal efficiency in the considered parameter space. The generated points are shown as blue dots.} 
\label{fig:2Deff_10}
\end{figure}

The signal efficiency for the selections described above is parameterized for each considered HNL mass as a function of $\log_{10}(c\tau)$, based on the generated grid of signal points. The results are presented in Figure~\ref{fig:2Deff_10}, where the signal points in the parameter space are shown as blue dots.
A very high efficiency, of order 80\%, is obtained for the range of couplings for which the HNL mean decay length is comparable with the size of the inner detector. Since the adopted parametrisation has large fluctuations for low efficiency values, the efficiency is conservatively set to zero when its parametrised value falls below 10\%.

\begin{figure}[h!]%
\centering
\includegraphics[width=0.7\textwidth]{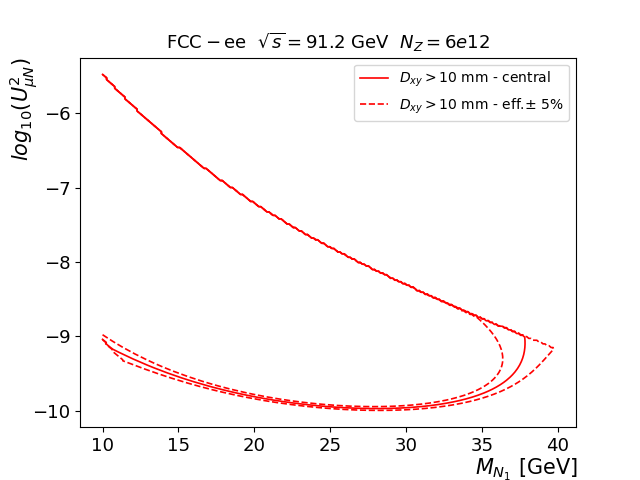}
\caption{$95\%$ CL exclusion limits in the parameter space. The solid line represents the central limit while the dashed lines correspond to shifts in the signal efficiency by $\pm 0.05$. The effect of the threshold set at 0.1 for the signal efficiency determines the behaviour of the upper part of the exclusion curve from the corner at high $M_{N_1}$ towards higher couplings.} 
\label{fig-hnl_limits_10}
\end{figure}

The final sensitivity for the HNL signal has been estimated at $95\%$ CL assuming no SM background survives the final selection. The number of expected signal events, $N_{exp}$, is normalised to the production of $6\times10^{12}$ $Z$ bosons.
Assuming negligible background and no observed events, the $95\%$ CL limits on the number of signal events is $N_{95}=3$. Figure~\ref{fig-hnl_limits_10} shows $95\%$ CL limit contours. The solid line represents the central limit, while the dashed lines correspond to relative shifts in signal efficiency by $\pm 0.05$, accounting for the uncertainty associated with the parametrization model. The threshold set at 0.1 for the signal efficiency induces a visible feature in the upper part of the exclusion curve, starting from the corner at high $M_{N_1}$ and extending towards higher couplings.

Couplings down to $U^2_{\mu N}\sim 10^{-10}$ for a range of $N_{1}$ masses between 20 and 35 GeV can be excluded by this analysis.

%% file: mujj_reco.tex
The semi-leptonic decay $N_1\rightarrow\mu jj$ proceeds through a virtual $W$, and has a branching fraction of $\sim50\%$. Compared to all other HNL decays it has the advantage that it does not contain neutrinos, which allows for the full reconstruction of the decay and also of the neutrino recoiling against the HNL  using the beam constraints. Due to this, it is an ideal channel for the study of prompt HNL production, which suffers from large SM backgrounds, which can be reduced with appropriate kinematic cuts.
The main backgrounds originate from:
\begin{itemize}
    \item hadronic decays of the $Z$ boson with a non-isolated muon in the final state from the decay of a meson inside the hadronic jet;
    \item events where Z decays into $\tau$  pairs, which may contain a muon, jets, and a high missing momentum;
    \item the four-fermion process $e^+e^-\to \mu \nu qq'$ which is an irreducible background.
\end{itemize}
A long-lived analysis is also possible, exploiting the precise vertex reconstruction provided by the high track multiplicity of the jets and the high branching ratio.

\subsection{Event reconstruction}
The final state of interest consists of two hadronic jets and a muon, all produced  at a single vertex with a distance from the undetected interaction point
corresponding to the flight path of the HNL in the tracking detector.
Thus, the reconstruction of the event kinematics and signal/background
separation rely on the jet reconstruction and vertexing capability 
of the IDEA detector. The relevant reconstruction algorithms are briefly 
discussed below.

The jets are reconstructed through a two-step procedure. They are first reconstructed
with the inclusive Durham $k_T$ algorithm with a 5 $\egev$ cut on the merging scale using the \verb+FASTJET+ package. If the algorithm returns one or two jets, these are used for the analysis, and if a larger number of jets is produced,
the exclusive $k_T$ algorithm is run requesting exactly two jets.

\begin{figure}[h!]%
\centering
\includegraphics[width=0.32\textwidth]{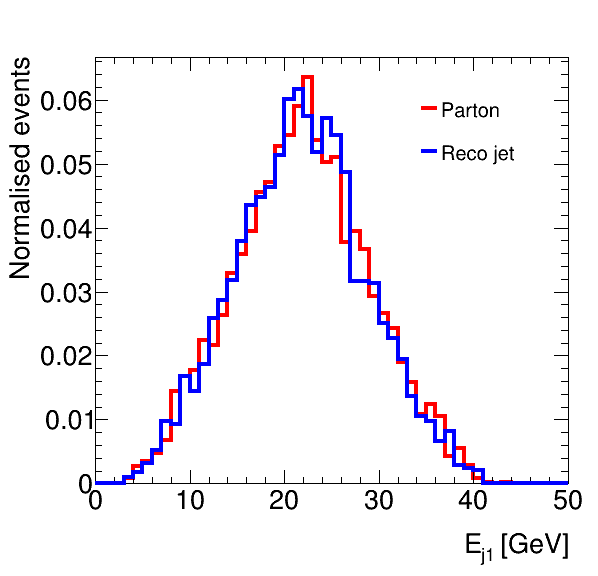}
\includegraphics[width=0.32\textwidth]{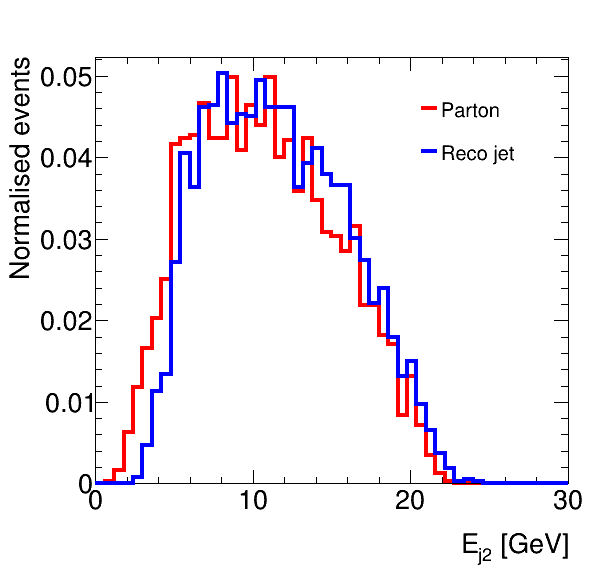}
\includegraphics[width=0.32\textwidth]{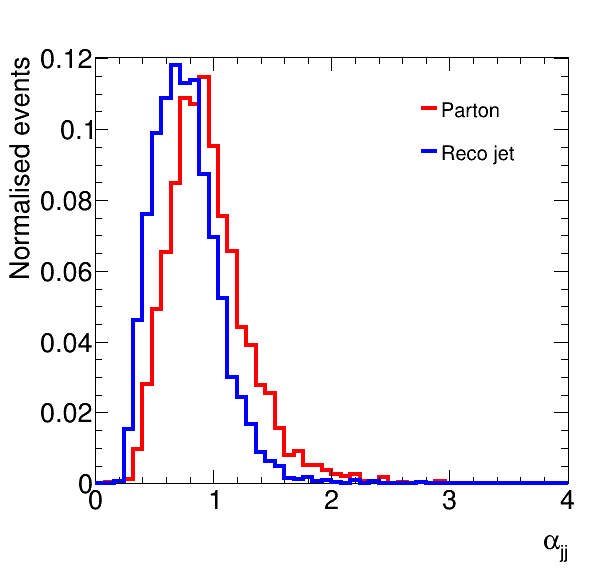}
\caption{Energy and angular distributions of jets compared to those of the partons. Durham $k_T$ algorithm, $\mhn=20\mgev$.\label{fig:jet20}}
\end{figure}

\begin{figure}[h!]%
\centering
\includegraphics[width=0.32\textwidth]{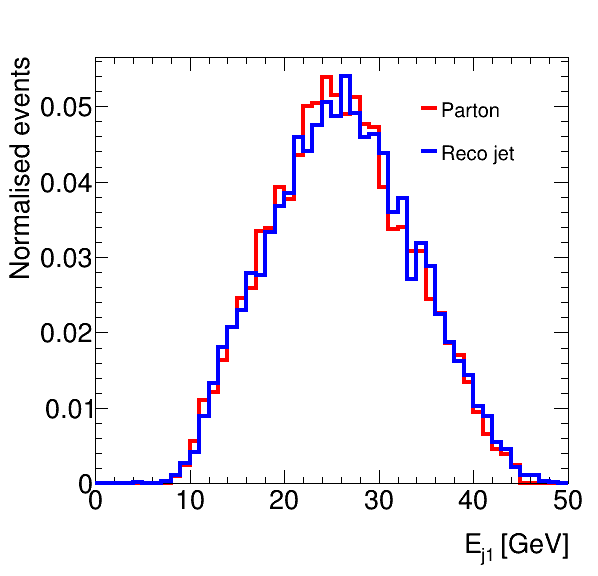}
\includegraphics[width=0.32\textwidth]{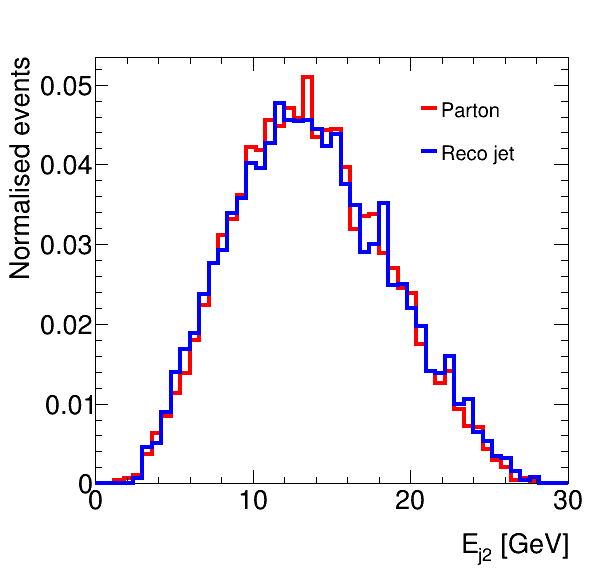}
\includegraphics[width=0.32\textwidth]{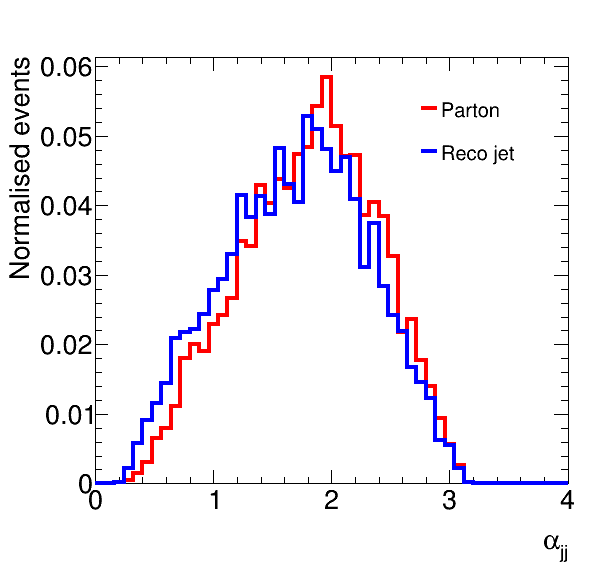}
\caption{Energy and angular distributions of jets compared to those of the partons. Durham $k_T$ algorithm, $\mhn=50\mgev$.}\label{fig:jet50}
\end{figure}

This algorithm yields a reasonable match 
between the kinematics of the reconstructed jets and  the originating partons  for all the HNL masses considered. This is  shown in Figures~\ref{fig:jet20} and 
\ref{fig:jet50}, where the energy distributions and the angular separation of the two jets are compared to the equivalent partonic quantities 
for two different values of the HNL mass, 20 and 50 GeV. As expected, for  low HNL masses, when the two jets are collimated, the algorithm somewhat underestimates
the jet-jet separation.

\begin{figure}[]
\centering
\includegraphics[width=0.45\textwidth]{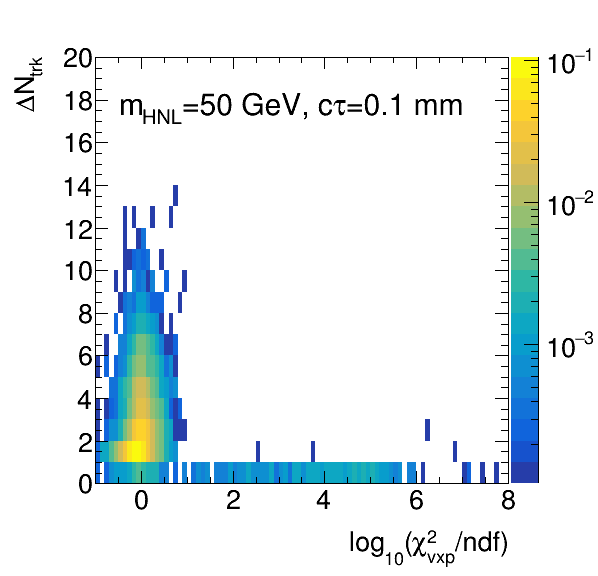}
\includegraphics[width=0.45\textwidth]{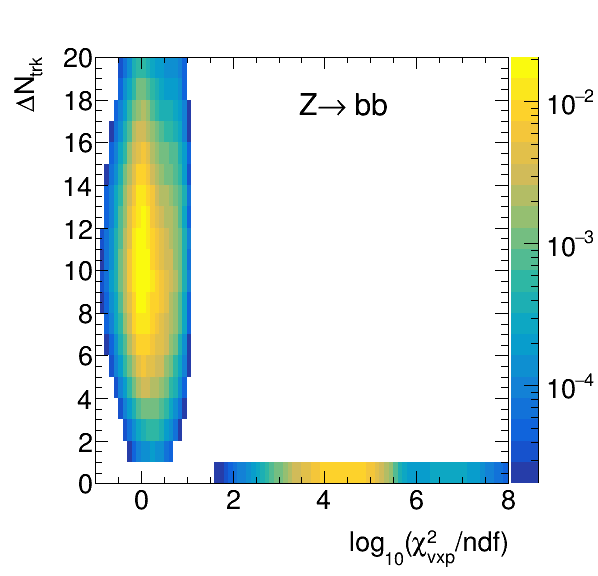}
\caption{Correlation of the difference between the total number of reconstructed
tracks in the event and the tracks attached to the primary vertex with
the value of $\log_{10}(\chi^2_{vx}/\mathrm{ndf})$
for an example 50 GeV
signal (Left) and $Z\rightarrow bb$ (Right) after the selection cuts.
} \label{fig:mujjvtx2d}
\end{figure}

The initial step of vertex reconstruction builds a vertex ($vx$) out of all 
the reconstructed tracks in the events, as described in Section \ref{sec:simulation}. 
The next step is the identification of a good `primary vertex' ($vxp$) defined as follows. The vertexing algorithm iteratively removes tracks with the largest contribution to the vertex $\chi^2/\mathrm{ndf}$ up to the point when the $\chi^2/\mathrm{ndf}$ value is stable. The vertex thus reconstructed is taken as $vxp$.
Two useful quality parameters for the event  are $\chi^2_{vxp}/\mathrm{ndf}$,  and the difference between the total number of tracks and those attached to the primary vertex, $\Delta N_{trk}\equiv N_{trk}^{vx} - N_{trk}^{vxp}$. 
The two -dimensional distributions of $\Delta N_{trk}$ versus 
$\log_{10}(\chi^2_{vxp}/\mathrm{ndf})$ for signal and $Z\to b\bar{b}$ 
events are shown in Figure \ref{fig:mujjvtx2d}.
For the signal, there is a small population of badly reconstructed vertices at very high $\chi^2_{vxp}/\mathrm{ndf}$,
and the good vertices have $\log_{10}(\chi^2_{vxp}/\mathrm{ndf})<1$; $\Delta N_{trk}$ is much larger
for $Z\to b\bar{b}$, corresponding to events where the two $b$-quarks have different decay paths before decaying, thus creating two or three separate vertices, if one considers the fragmentation tracks originating from the $Z$ decay vertex. 
Selections on these variables, as well as on the position of the primary vertex, 
will be used in the following both for event cleaning purposes and for reducing the heavy-flavour background.

%% file: mujj_ana.tex
\subsection{Preselection}
The analysis starts with a filtering step, needed to bring the large background samples of $Z$ decay down to a manageable size. The filter requirements are at least a reconstructed muon with energy greater than
3 $\egev$, missing energy larger than 5 $\egev$ and at least three reconstructed tracks.
The final state of interest is then selected by requiring only one reconstructed muon, at least one jet with energy in excess of 3 GeV, and vetoing reconstructed electrons.
A good primary vertex inside the tracking detector is required, with $\log_{10}(\chi^2_{vxp}/\mathrm{ndf})<1$, and $\Delta N_{trk}<11$, which has approximately full efficiency for well-reconstructed signal events, as shown in Figure~\ref{fig:mujjvtx2d}.
The event is also required to be well contained in the detector, by imposing $|\cos{\theta_{m}}|<0.98$ where $\theta_{m}$ is the angular distance of the missing energy vector from the beam direction.  The muons in the $Z$ backgrounds are predominantly produced alongside neutrinos in the decay of heavy quarks and taus, and are thus well aligned with the missing energy vector. An additional requirement $\cos(\alpha_{lm})<0.99$ is applied,
where $\alpha_{lm}$ is the angle between the missing energy and the muon. 
This ensemble of requirements is called the preselection, and is applied to all events.

At this point two different analyses are developed, an analysis targeting `prompt' decays of the HNL, and an analysis targeting long-lived decays. The separation of the two analyses is to some extent arbitrary. We use the variable $\rvxp$ for the separation, the
radial distance of the primary vertex from the centre of the detector, and we define the separation point at $\rvxp=0.5~\mathrm{mm}$. Two different strategies are adopted. 
For the prompt analysis a significant background will always be present, and a standard kinematic analysis optimising the significance of signal over background is performed. 
For the `long-lived' analysis, after the $\rvxp>0.5~\mathrm{mm}$ cut there is a relatively small residual background, which can be fully removed by kinematic selections which retain most of the long-lived HNL signal.
\subsection{Prompt analysis}
The final state topology of signal events depends on the HNL mass. The angle between the two jets of the HNL decay decreases as the HNL mass decreases, as one can see by comparing the rightmost plot in Figures~\ref{fig:jet20} and \ref{fig:jet50}, and there is an increasing probability that the two jets are reconstructed as a single jet.
Thus, for lower HNL mass the signal is more likely to display a one-jet signature, while at higher masses the two-jet signature dominates. The percentage of events with a single cluster ranges from $\sim 55\%$ for $\mhn=20\mgev$ to $\sim 6\%$ for $\mhn=70\mgev$. 

Two different analysis strategies are implemented for the 1-jet and the 2-jet cases, chosen to have a higher signal yield at low (high) \mhn respectively.  

\begin{figure}[h!]%
\centering
\includegraphics[width=0.49\textwidth]{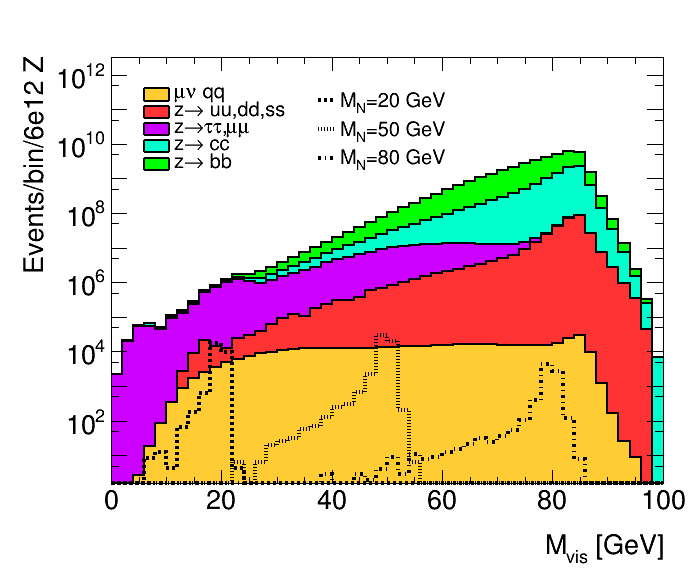}
\includegraphics[width=0.49\textwidth]{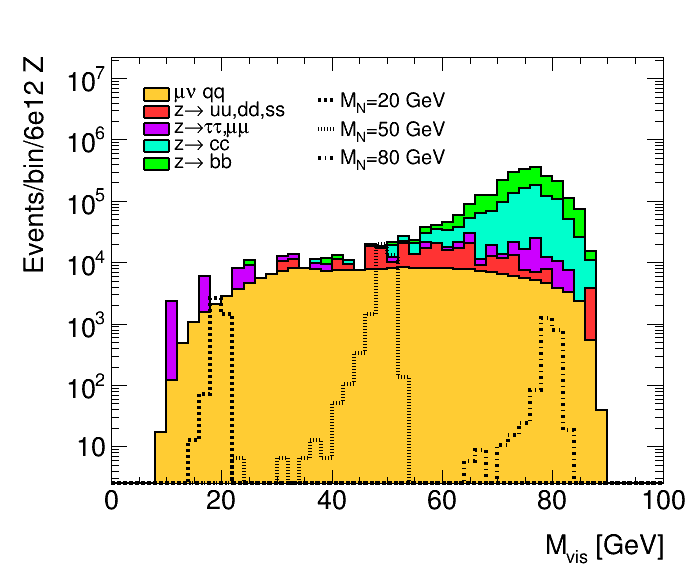}
\caption{Distribution of the reconstructed visible mass 
for the backgrounds and three signal samples. Plot on the left: after preselection. Plot on the right: after kinematic selections as defined in Table~\ref{tab:selection}. The events are  normalised to the FCC-ee $Z$-pole run statistics.}
\label{fig:mujjmvis}
\end{figure}

The selection strategy is determined by the nature of the dominant $Z$ decay background in the different HNL mass ranges. This can be seen in the left panel of Figure~\ref{fig:mujjmvis}, where the distribution of the visible invariant mass $M_{vis}$ built with the jet(s) and the muon is shown for three signal examples and for the stacked backgrounds after preselection. Both 1-jet and 2-jet events enter the distributions. 
The irreducible 4-fermion background is approximately independent of the mass and always subdominant. For  $\mhn<30\mgev$ $Z\rightarrow\tau\tau$ is dominant, whereas for $\mhn>30\mgev$ all the $Z$ decay channels contribute, dominated by the decays into $b$ and $c$-quarks in the higher part of the mass range. 

\begin{figure}[h!]%
\centering
\includegraphics[width=0.49\textwidth]{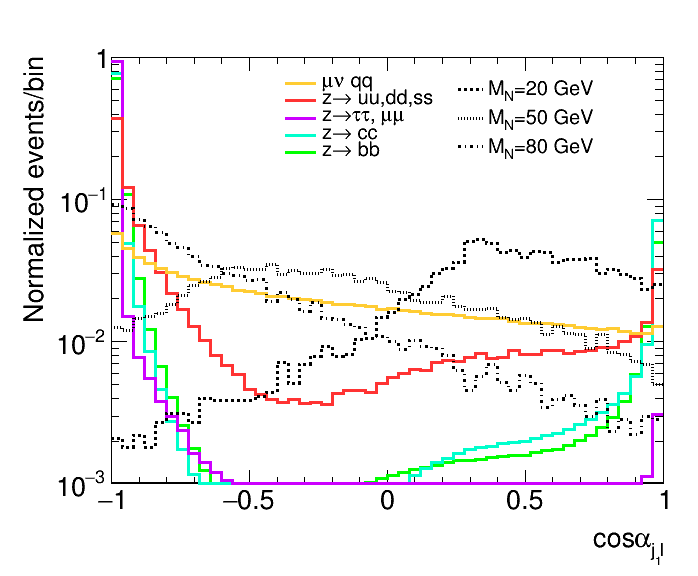}
\includegraphics[width=0.49\textwidth]{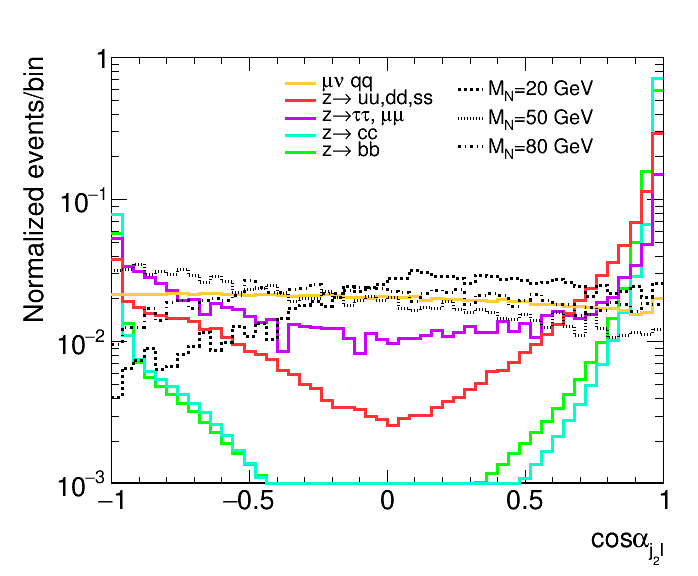}
\includegraphics[width=0.49\textwidth]{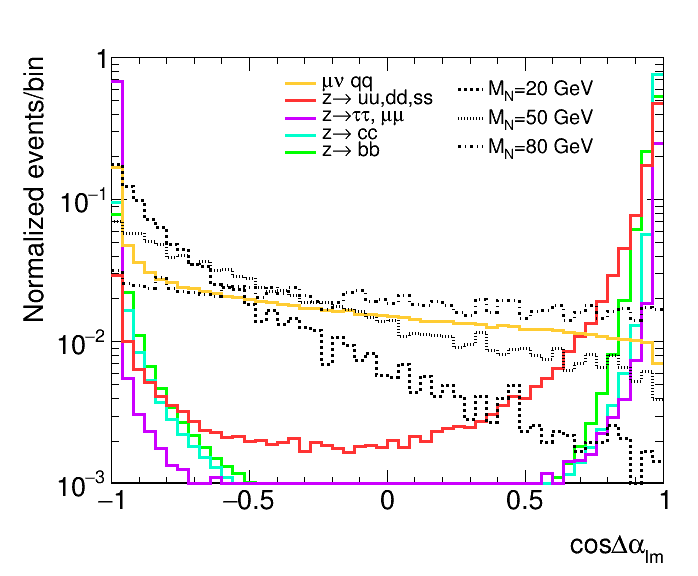}
\includegraphics[width=0.49\textwidth]{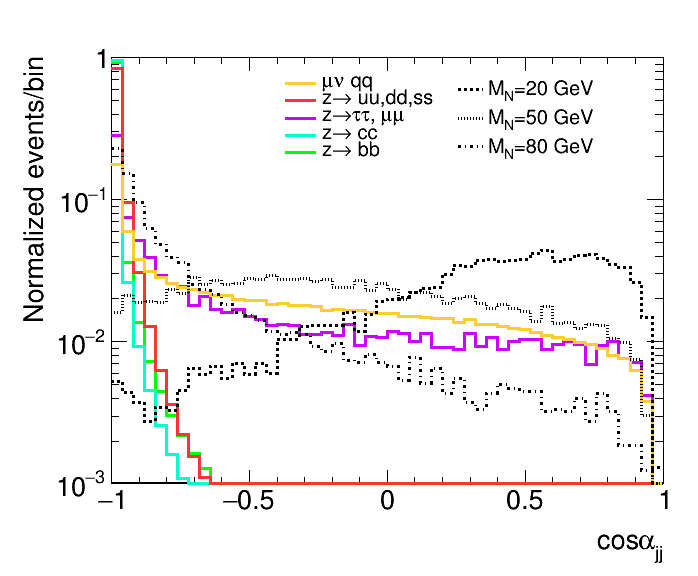}
\caption{Normalized distributions of the angular discriminant variables discussed in the 
text for the backgrounds and three signal samples. The distribution are after preselection and for the prompt analysis. For the left column no selection is applied  on the number of jets, the plots in the right column are for events with two reconstructed jets.}
\label{fig:mujjvar1}
\end{figure}

For the $Z$ decay into quarks the dominant topology includes two approximately back-to-back jets with a muon very near one of the jets, and the missing energy 
aligned with the lepton. The discriminant variables resulting from this observation 
are the cosines of the angular distances between the muon and each of the two jets,  $\alpha_{j_1l}$ and $\alpha_{j_2l}$, shown in the upper row of Figure~\ref{fig:mujjvar1}, and of the angle $\alpha_{lm}$ between the lepton and the missing momentum, shown in the lower left panel.
For the two-jet case, the angular difference between jets, $\alpha_{jj}$, is also a good discriminant, as shown in the lower right panel of the figure.

The presence in the event of two quarks with a long lifetime can be exploited by applying selections on $\Delta N_{trk}$ defined in the previous section, and on the transverse impact parameter of the muon with respect to the centre of the detector $D_{0,\mu}$.
For the higher range of masses considered, $\geq 70$~GeV, the requirement 
that $D_{0,\mu}$  be smaller than 8 times its resolution
is fully efficient for signal and provides additional rejection of heavy flavour backgrounds.

\begin{figure}[h!]%
\centering
\includegraphics[width=0.49\textwidth]{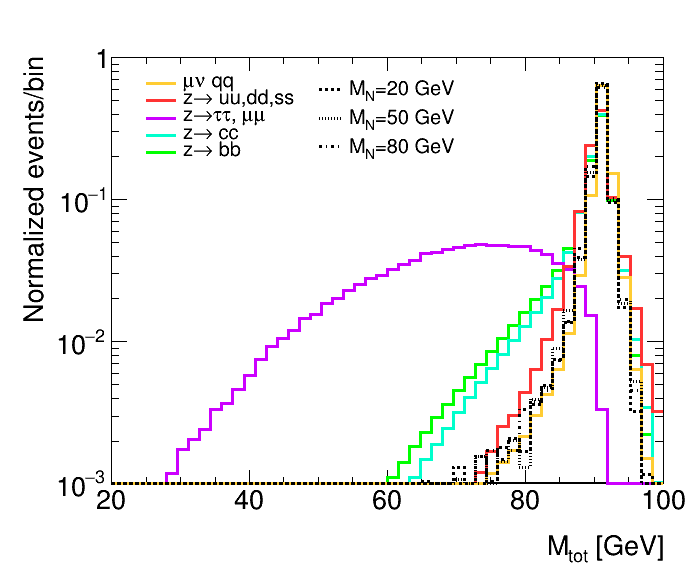}
\caption{Normalized distribution of the total reconstructed 
mass of the event for the backgrounds and three signal samples. The distribution are
after preselection and for the prompt analysis.}
\label{fig:mtot}
\end{figure}

For $Z\rightarrow\tau\tau$ the dominant topology is the case where on one leg the $\tau$ decays into a hadronic jet and neutrino, and on the other leg it decays into a muon and two neutrinos. The preferred topology would be a lepton back-to-back to a single jet, and a cut on the angle between them, $\alpha_{jl}$, would strongly reduce this background.  The signal has a single neutrino in the final state, 
and the invariant mass built from the visible particles and the missing energy 
($M_{tot}$) is peaked at the $Z$ mass. For the $Z\rightarrow\tau\tau$ with at least two neutrinos in the final state, $M_{tot}$ has a large tail towards lower masses. This is also true for part of the heavy flavour decays, as shown in Figure~\ref{fig:mtot}.

Based on these considerations, selection criteria based on the variables described above were applied to the events, where the cuts were optimised to achieve the best significance for the two topologies considered. The values
of the selections are given in detail in Table~\ref{tab:selection}.
\begin{table}
\centering
\begin{tabular}{|*{5}{c|}}  
\hline
\hline
\multicolumn{5}{|c|}{Preselection} \\
\hline
\hline
	Filter & \multicolumn{4}{|c|}{$N_{\mu}>0,\; N_{tracks}\ge3,\; E_m>5~\mathrm{GeV}$} \\ \hline
	Final state &   \multicolumn{4}{|c|}{$N_{\mu}=1,\; N_{ele}=0,\; N_{jets}>0$} \\ \hline
	Vertexing &    \multicolumn{4}{|c|}{$\rvxp<2~\mathrm{m},\; z_{vxp}<2~\mathrm{m},\; \mathrm{log_{10}}(\chi^2_{vxp}/\mathrm{ndf})<1 $,\; $\Delta N_{trk}<11$} \\ \hline
	Kinematics &    \multicolumn{4}{|c|}{$|\cos\theta_{m}|<0.98$, \; $\cos\alpha_{lm}<0.99$} \\
\hline
\hline
\hline
\multicolumn{5}{|c|}{Kinematic selections} \\
\hline
\hline
Variable  & \multicolumn{2}{|c}{Prompt} & \multicolumn{2}{|c|}{LLP} \\ \hline
\hline
$\rvxp$ [mm]   & \multicolumn{2}{|c}{$<0.5$} & \multicolumn{2}{|c|}{$>0.5$} \\ \hline
	$N_{jets}$ & \multicolumn{1}{|c}{1} & \multicolumn{1}{|c}{2} & \multicolumn{2}{|c|}{$>0$} \\
\hline
	$\mhn$ [GeV] & \multicolumn{2}{|c}{ - } & \multicolumn{1}{|c}{$\leq30$} & \multicolumn{1}{|c|}{$>30$} \\ \hline
\hline
	$M_{tot}$ [GeV] &  \multicolumn{2}{|c}{$>85$}  & \multicolumn{2}{|c|}{$>85$}\\ \hline
	$|\cos\theta_{m}|$ &  \multicolumn{2}{|c}{<0.95}  & \multicolumn{2}{|c|}{$<0.98$}\\ \hline
	$\cos\alpha_{jj}$ &  \multicolumn{1}{|c}{--} & \multicolumn{1}{|c}{$>-0.8$} &    \multicolumn{1}{|c}{--} & \multicolumn{1}{|c|}{$>-0.98$ (2j)}\\ \hline
	$\cos\alpha_{lm}$ &  \multicolumn{1}{|c}{<0.5} & \multicolumn{1}{|c}{$<0.8$} &            \multicolumn{1}{|c}{--} & \multicolumn{1}{|c|}{<0.9}\\ \hline
	$\cos\alpha_{j_1l}$ &  \multicolumn{1}{|c}{$\in [-0.5,\, 0.96]$} & \multicolumn{1}{|c}{$\in [-0.8,\, 0.98]$} &          \multicolumn{1}{|c}{$>-0.5$} & \multicolumn{1}{|c|}{ $\in [-0.99,\, 0.99]$}\\ \hline 
	$\cos\alpha_{j_2l}$ &  \multicolumn{1}{|c}{--} & \multicolumn{1}{|c}{$\in [-0.8,\, 0.98]$} &          \multicolumn{1}{|c}{--} & \multicolumn{1}{|c|}{$\in [-0.99,\, 0.99]$ (2j)}\\ \hline 
	$\Delta N_{trk}$ &  \multicolumn{2}{|c}{<5} &    \multicolumn{1}{|c}{--} & \multicolumn{1}{|c|}{$<5$}\\ \hline
	$D_{0,\mu}$ & \multicolumn{2}{|c}{$>8\sigma \, (\mhn>70~\mathrm{GeV})$}  & \multicolumn{2}{|c|}{--} \\ \hline
\hline
\hline
\multicolumn{5}{|c|}{Mass selection} \\
\hline
\hline
	$E_ {miss}$ [GeV] &  \multicolumn{2}{|c}{$\in E_{\nu} \pm 2 \times 10\% \sqrt{E_{\nu}}$} &  \multicolumn{1}{|c}{>38} &  \multicolumn{1}{|c|}{>20} \\
	$M_{vis}$ [GeV]  & \multicolumn{2}{|c}{$\in \mhn \pm 2\times 10\%  \sqrt{\mhn}$} &  \multicolumn{1}{|c}{<35} &  \multicolumn{1}{|c|}{--} \\
\hline
\end{tabular}
\caption{Selection criteria for the different branches of the $HNL\rightarrow\mu jj$ analysis}\label{tab:selection}
\end{table}

The incremental efficiencies for the signal, and the number 
of predicted background events surviving the different analysis stages are
shown in Tables~\ref{tab:sigeff_prompt} and \ref{tab:bgnum} respectively. 
\begin{table}
\centering
\begin{tabular}{|l|c|c|c|c|c|}
\hline
\hline
\multicolumn{6}{|c|}{Signal efficiencies} \\
\hline
\hline
	\mhn    & Filter & Preselection & Kinematic & Mass & Total\\
	$[\mathrm{GeV}]$   &        &      &  selection   &  selection              &       \\
\hline
10 & 0.77 & 0.62 & 0.57 & 0.92 & 0.25 \\
20 & 0.88 & 0.77 & 0.57 & 0.87 & 0.33 \\
30 & 0.94 & 0.84 & 0.65 & 0.89 & 0.45 \\
40 & 0.96 & 0.88 & 0.67 & 0.89 & 0.50 \\
50 & 0.97 & 0.89 & 0.63 & 0.87 & 0.47 \\
60 & 0.97 & 0.90 & 0.55 & 0.87 & 0.42 \\
65 & 0.97 & 0.91 & 0.50 & 0.85 & 0.38 \\
70 & 0.97 & 0.91 & 0.45 & 0.85 & 0.34 \\
80 & 0.91 & 0.90 & 0.28 & 0.81 & 0.18 \\
85 & 0.40 & 0.91 & 0.09 & 0.75 & 0.03 \\
\hline
\end{tabular}
\caption{Incremental efficiencies of the different stages of the event selection
	as a function of \mhn for prompt signal.}\label{tab:sigeff_prompt}
\end{table}
\begin{table}
\centering
\begin{tabular}{|l|c|c|c|}
\hline
Sample & Produced & Preselection & Selection \\
       &          & $\rvxp<0.5$~mm & $\rvxp<0.5$~mm \\
\hline
$Z\rightarrow b\bar{b}$ & 9.36e+11 &  2.98e+10 & 5.57e+05\\
$Z\rightarrow c\bar{c}$ & 6.96e+11 &  1.08e+10 & 4.76e+05\\
$Z\rightarrow s\bar{s}$ & 9.36e+11 &  1.43e+08 & 7.68e+04\\
$Z\rightarrow u\bar{u},\, d\bar{d}$ & 1.60e+12 &  2.32e+08 & 2.29e+04\\
$Z\rightarrow\mu^+\mu^-$ & 2.20e+11 &  1.21e+06 & 6.59e+03\\
$Z\rightarrow\tau^+\tau^-$ & 2.20e+11 &  1.38e+08 & 4.46e+05\\
$\mu\nu qq^{\prime}$ & 6.56e+05 &  4.91e+05 & 2.21e+05\\
\hline
\end{tabular}
\caption{Background events passing each stage of the selection normalised to the 
expected statistics of the FCC-ee $Z$-pole run}\label{tab:bgnum}
\end{table}

\begin{figure}[h!]%
\centering
\includegraphics[width=0.49\textwidth]{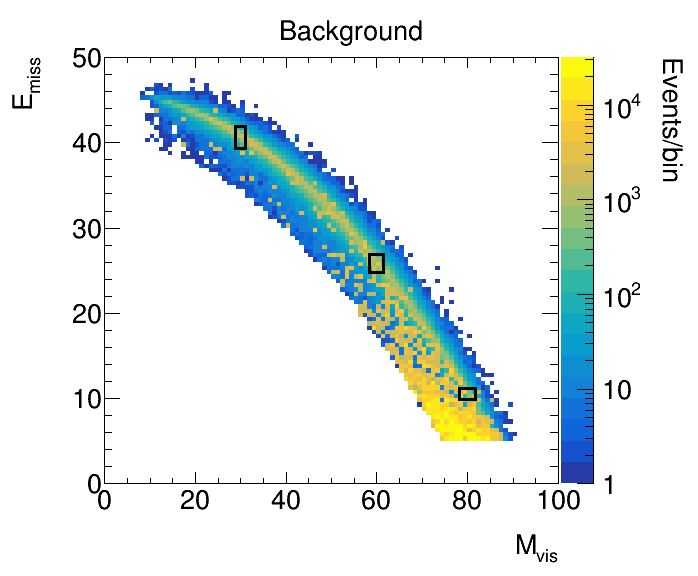}
\includegraphics[width=0.49\textwidth]{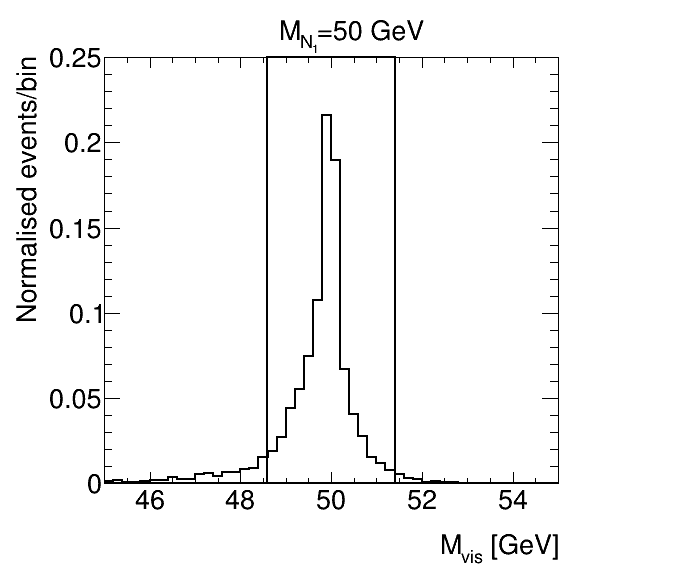}
	\caption{Left: Distribution of $E_{miss}$ versus $M_{vis}$ for the 
backgrounds after selection cuts. The black rectangles show the effect of
the mass selection for three values of the HNL test mass: 30, 60 and 80 GeV. Right:  Distribution of the reconstructed visible mass for a signal point at $\mhn=50\egev$.} 
\label{fig:mvis1}
\end{figure}

After these cuts  a significant amount of backgrounds is still present, especially at high HNL masses, as shown in the right panel of Figure \ref{fig:mujjmvis}. Additional rejection can be achieved by considering a grid of test values for \mhn.  
For each HNL test mass \mhn, the $M_{vis}$ variable has a peak around \mhn, as shown in Figure~\ref{fig:mujjmvis}. 
Similarly for a given value of \mhn and $e^+e^-$ collisions at the $Z$-pole the energy of the neutrino recoiling against the HNL, $E_{\nu}(\mhn)$, has a fixed value of:
\begin{equation}
E_{\nu}(\mhn) = \frac{M_{Z}^2 - \mhn^2}{2\, M_{Z}}.
\end{equation}
The correlation between the two variables for the background is shown in 
the left panel of Figure~\ref{fig:mvis1}.

The resolution on $M_{vis}$ obtained with the \verb+DELPHES+ simulation is shown in the right panel of Figure \ref{fig:mvis1} for a signal with $\mhn=50\mgev$. The distribution is not gaussian, as the jets are reconstructed in \verb+DELPHES+ with an idealised version of the particle flow algorithm which does not account for uncertainties on the amount of calorimetric energy associated with a track.  It is found that about 90\% of the signal is contained within a mass window centred around the nominal visible mass with a width scaling approximately as $2\times 10\% \times \sqrt{M/ \mgev}$. 
A window in the ($M_{vis}$, $E_{miss}$) plane defined as:
\[
\begin{split}
M_{vis} \in \mhn \pm 2\times 10\%  \sqrt{\mhn/\egev} \\
E_{miss} \in E_{\nu}(\mhn) \pm 2 \times 10\% \sqrt{E_{\nu}/ \egev} \\
\end{split},
\]
shown in the left side of Figure~\ref{fig:mvis1} for three values of \mhn, contains between 80 and  90\% of the signal, except for the highest values of $\mhn$, and selects only a small fraction of the background.

\begin{figure}[h!]%
\centering
	\includegraphics[width=0.8\textwidth]{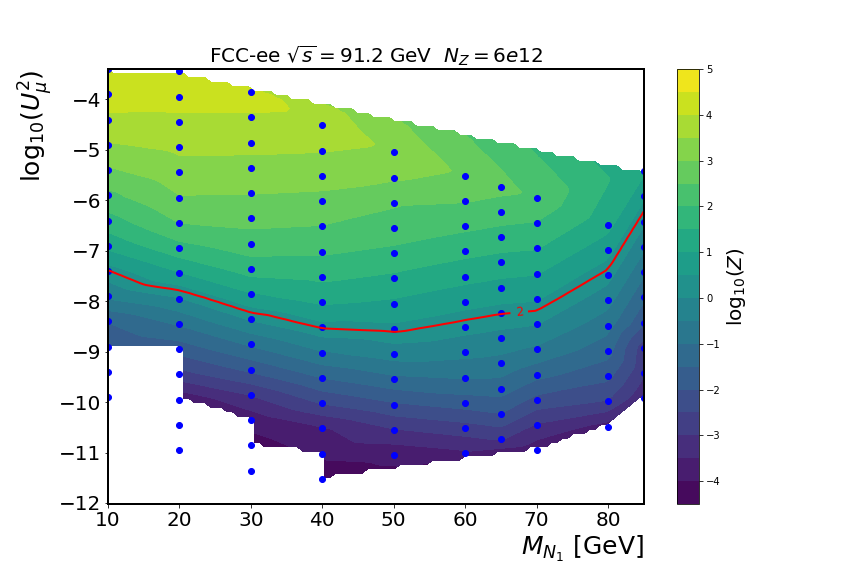}
	\caption{Map of the sensitivity $z$ of the prompt analysis 
        $\mhn-\log_{10}(U_{\PGm}^2)$ plane. The generated points are shown as blue dots,
        and the line corresponding to $z=2$ is shown in red.}
\label{fig:promptsigni}
\end{figure}

After mass selection, for each point in the generated signal grid the numbers of signal and background events are calculated normalised to the expected statistics of the FCC-ee $Z$-pole run. The statistical significance $z$ for each point is calculated based on the prescription in~\cite{ATLASstat} and is shown in Figure~\ref{fig:promptsigni}. The line corresponding to $z=2$, the 95\% CL exclusion for the signal was calculated by interpolating through the generated points, which are shown as blue dots in the plot. 

\subsection{Long lived analysis}
As discussed above, long-lived signal regions are defined by the requirement $\rvxp>0.5$~mm. This selection removes the irreducible 4-fermion background.
The signal has a significant long-lived cross-section for $\mhn<70~\mathrm{GeV}$,  a region in which $Z\rightarrow\tau\tau$ is the dominant background, and the contribution from the decay of $Z$ into heavy quarks is strongly reduced. Therefore, as for the fully leptonic decays, we define a selection which reduces the backgrounds to zero, while maximising signal efficiency, and we define the 95\% sensitivity area as the area where at least three signal events survive the selections for the expected integrated luminosity of the $Z$-pole run of the FCC-ee.

A selection on $M_{tot}>85~\mathrm{GeV}$, ensuring that no neutrino is produced, strongly reduces $Z\rightarrow\tau\tau$. 
A veto on the topology where the muon and the leading jet are approximately back to back, and in the two-jet case where the jets are back to back, together with the request that $E_{miss}>38~\mathrm{GeV}$ fully eliminate the background for values of the HNL test mass smaller than or equal to $30~\mathrm{GeV}$.

\begin{figure}[h!]%
\centering
        \includegraphics[width=0.49\textwidth]{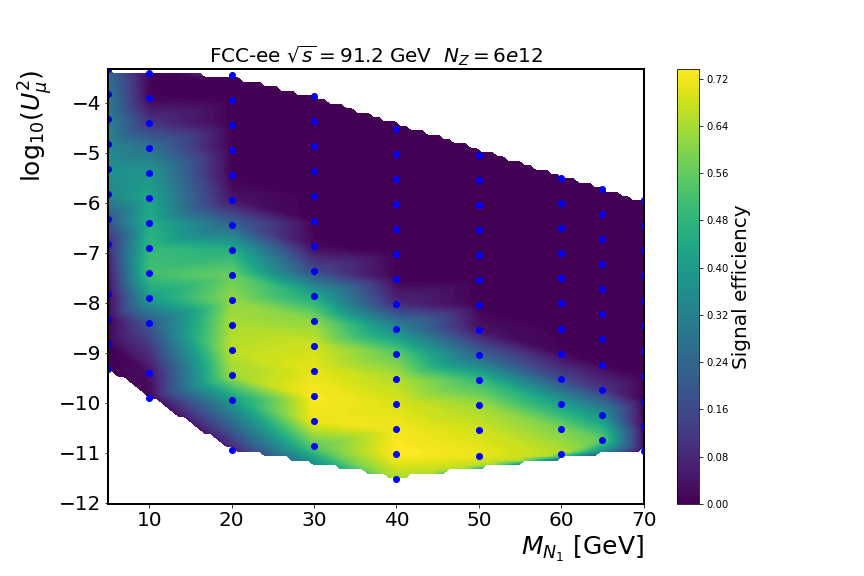}
        \includegraphics[width=0.49\textwidth]{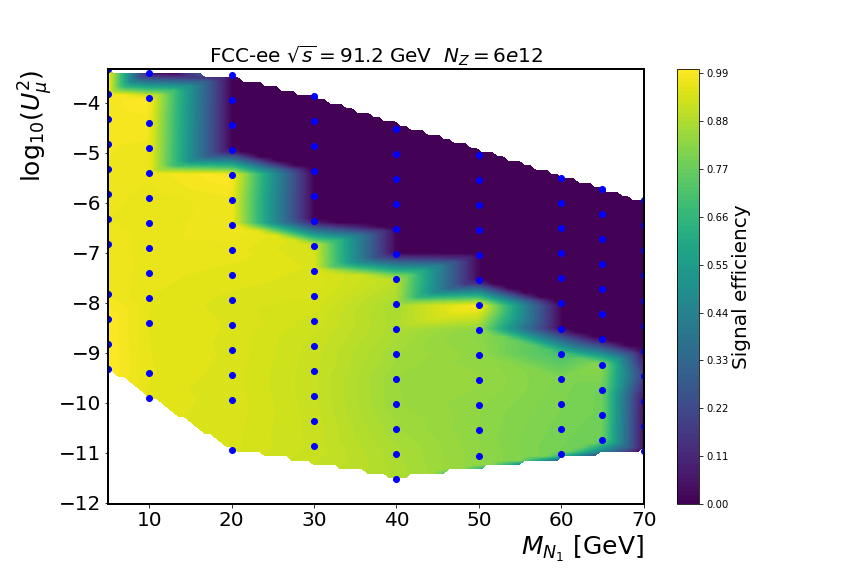}
        \caption{Efficiency of the long-lived selections in the
        $\mhn-\log_{10}(U_{\PGm}^2)$ plane. On the left: total selection efficiency;
	on the right: efficiency of the kinematic cuts with respect to the events
	with a reconstructed vertex in the inner tracker.
	The generated points are shown as blue dots.}
\label{fig:llpeffi}
\end{figure}

For higher HNL test masses, in the range 30-65~GeV, the additional contributions from heavy quarks demand more complex selection criteria, based on the same angular variables as for the prompt analysis.
These criteria are shown in Table ~\ref{tab:selection}, and reduce the background to zero, while retaining high signal efficiency.
The values of the selection efficiencies in the $\mhn-\log_{10}(U_{\PGm}^2)$ plane are shown in Figure~\ref{fig:llpeffi}. On the left panel the total selection efficiency is displayed, incorporating both the request of a well-reconstructed vertex and the kinematic selection criteria; on the right panel we show the incremental efficiency of the kinematic selections. The latter is very high, between 75\% and almost 100\%, and for lower masses the experimental efficiency is completely dominated by the requirement that the decay occurs in the volume of the tracking detector and is reconstructed by the tracker.

As explained in Section~\ref{sec:simulation}, for the $Z$ decay backgrounds  only $\sim3\times10^{9}$ Monte Carlo events were generated, well short of the expected statistics at the FCC-ee $Z$-pole run. 
In order to have some confidence that the zero-event condition can be reached, the number of $Z$ decay events after selections as a function of $\rvxp$ was fitted with an exponential for both signal regions defined above.
It was found that for a cut $\rvxp>0.1$~mm the extrapolated curve yields a prediction of less than one event in both cases, thus ensuring that no background from well-reconstructed events is left over. 
In addition, no $Z$ Monte Carlo events would pass the selection $\rvxp>0.2$~mm, and the reach curve was calculated for $\rvxp>1,\, 5\, \mathrm{and}\; 10$~mm, in addition to the nominal $\rvxp>0.5$~mm selection, to evaluate how much tighter cuts on the HNL decay length would affect the experimental reach.
\begin{figure}[h!]%
\centering
        \includegraphics[width=0.65\textwidth]{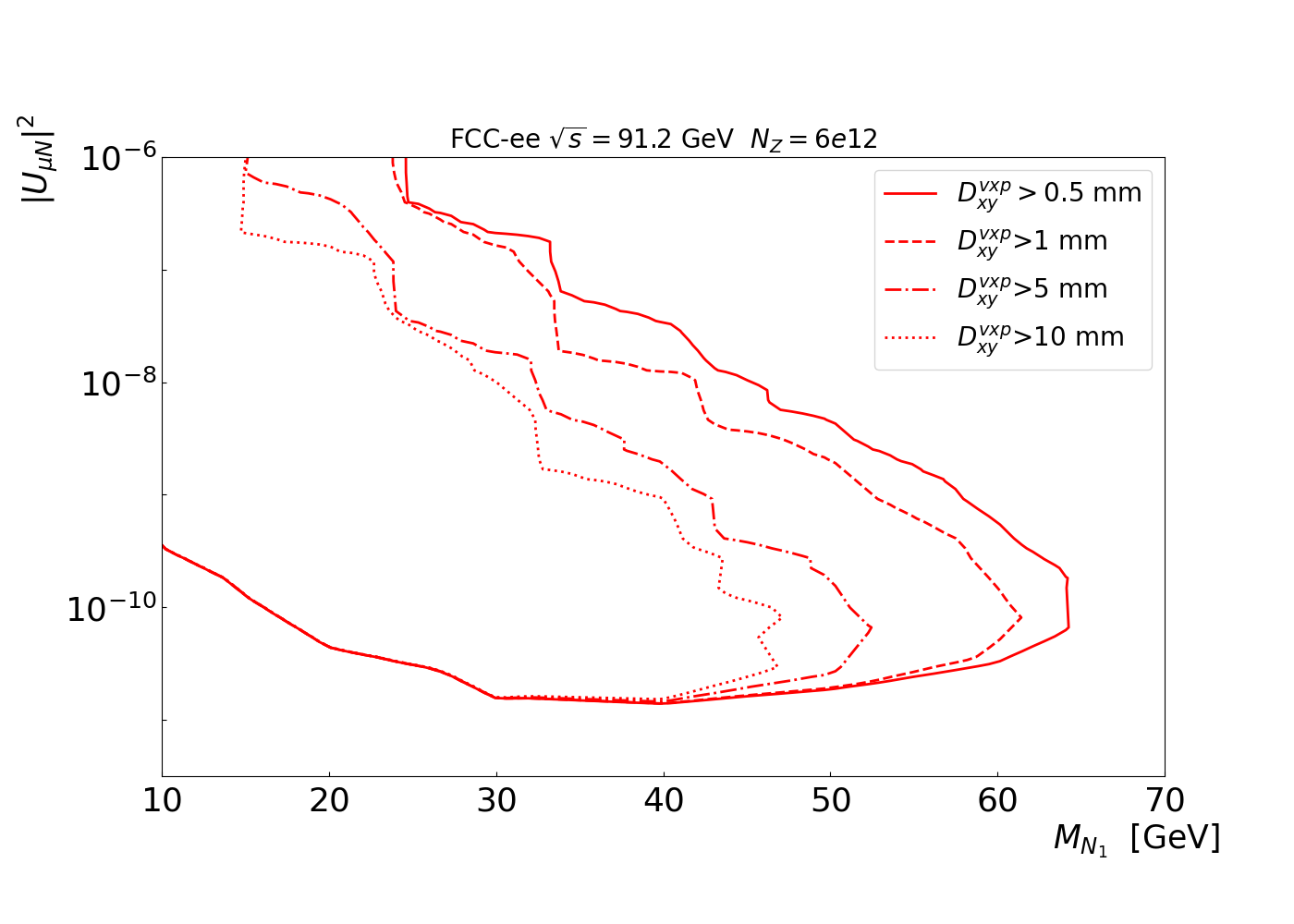}
        
        \caption{Curves bounding the areas where three events survive the long-lived
	selections in the $\mhn-\log_{10}(U_{\PGm}^2)$ plane for four values 
	of the threshold on the transverse position of the reconstructed 
	decay vertex $\rvxp$.} 
\label{fig:llpsigni}
\end{figure}

In Figure \ref{fig:llpsigni} we show the lines bounding the area with at least three detected events after selections in the $\mhn-\log_{10}(U_{\PGm}^2)$ plane corresponding to the 95\% CL experimental sensitivity. The results are given for the four different requirements on $\rvxp$ given above. The extension of the sensitivity region towards low values of $U_{\PGm}^2$ is little affected by an increase in the threshold set on $\rvxp$. The main effect would be a reduction of the accessible mass range towards high mass values. The loss in coverage towards higher $U_{\PGm}^2$ values would be  compensated by a corresponding increase of the coverage of the prompt analysis.

%% file: combined.tex
We have studied the experimental reach of the FCC-ee $Z$-pole run to the production
of an HNL in a simplified benchmark model with a single experimentally accessible HNL mixing only with muons. 

Two different decay channels for the HNL were addressed, a semileptonic one into
a muon and two jets, and a fully leptonic one into a $\mu^+\mu^-$ pair and a neutrino.
For the first case two analyses were performed, one addressing the `prompt'
decay of the HNL, and one addressing long-lived HNLs decaying after a measurable flight 
path in the detector; for the second channel only the long-lived  signature was studied.  The 95\% CL coverages of the analyses in the $\mhn-U_{\PGm}^2$ plane
were obtained based on an analysis of simulated events. It is found that  the long-lived analyses, thanks to the large $Z$ statistics allow very low values of the mixing to be reached, down to $U_{\PGm}^2\sim 10^{-11}$ for \mhn  between 5 and $\sim 65$~GeV. The `prompt' analysis has larger backgrounds, but it is useful for covering higher values of $U_{\PGm}^2$ and  \mhn up to 85~GeV.

\begin{figure}[h] \centering
\includegraphics[width=0.49\textwidth]{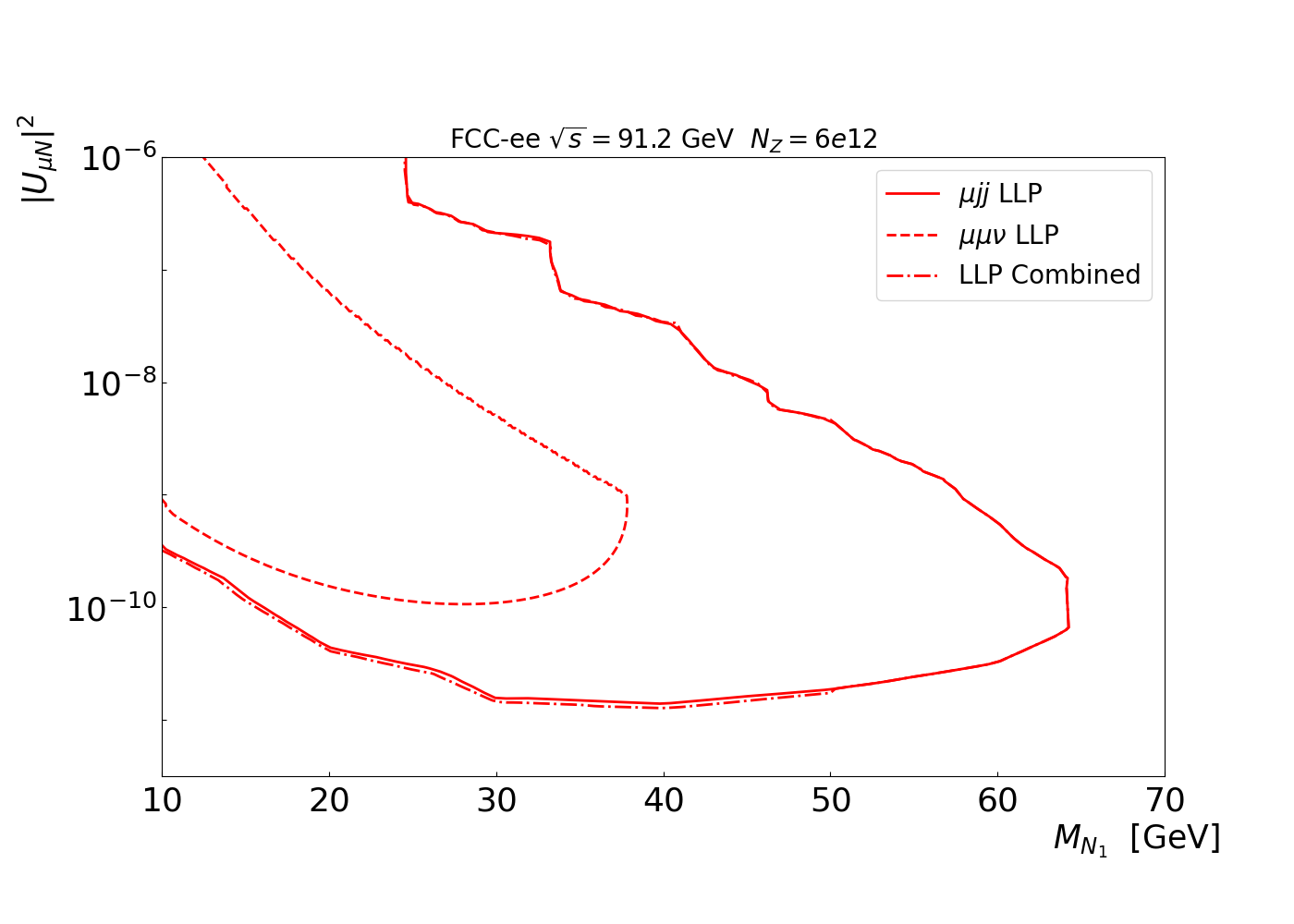}
\includegraphics[width=0.49\textwidth]{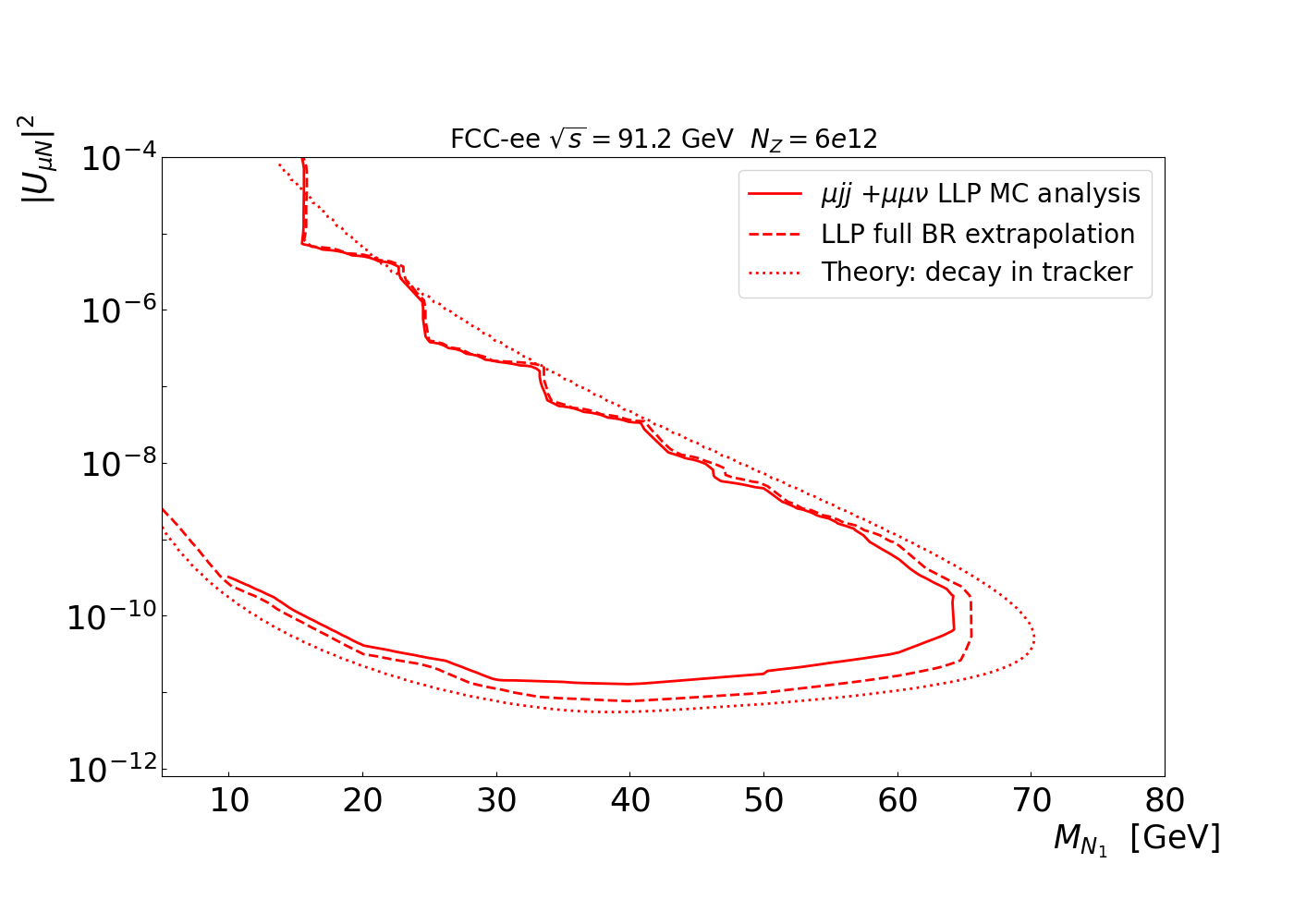}
\caption{Sensitivity limits at 95\% confidence level in the $\mhn-U_{\PGm}^2$ plane
	for the LLP analyses described in the text and for their statistical combination. Left panel: combination of the semileptonic and leptonic LLP analyses. Right panel: comparison of the coverage of the combined LLP analyses (full line) with the extrapolation to the full visible branching fraction of the HNL (dashed line), and with the line corresponding to three events with decay in the tracker (dotted line).}\label{fig:llpcomb}
\end{figure}

The statistical combination of the two LLP channels was performed by summing 
the number of expected signal events after the selections described above 
for each of the two channels on a grid in the $\mhn-U_{\PGm}^2$ plane.
The 95\% sensitivity region was calculated by interpolation as the area in which at least three signal events survive the selection,
and is shown as a dashed dotted line in the left panel of Figure~\ref{fig:llpcomb}, whereas the  semileptonic channel is shown as a full line, and the fully leptonic one as
a dashed line.
The combined analysis covers yields a gain of $\sim10\%$ 
in terms of covered values of $U_{\PGm}^2$ with respect to the semileptonic analysis alone, corresponding to the ratio of the branching fractions of the two considered channels. The present study is based on two Monte Carlo analyses that cover between 55 and 60\% of the visible HNL branching fractions. Channels which were not studied are dominated by the decay of the HNL into a neutrino and two jets via a virtual $Z$, and fully leptonic decays with a muon, a neutrino and an electron or a tau, via a virtual $W$. Under the assumption that the analysis efficiency will be the same as for the channels explicitly studied, the parameter space coverage for the combined analysis of all visible HNL decay channels can be calculated. The results are shown in the right panel of Figure~\ref{fig:llpcomb}, where the combined results of the two analyses of this paper are compared to the extrapolated reach for the complete visible decay channels and to the theoretical calculation of the line corresponding to three events decaying in the IDEA inner tracker based on the formulas of \cite{Drewes:2022rsk}.  With the analysis of all decay channels, mixing values as small as approximately 70\% of the theoretical limit could be covered.

\begin{figure}[h] \centering 
\includegraphics[width=0.9\textwidth]{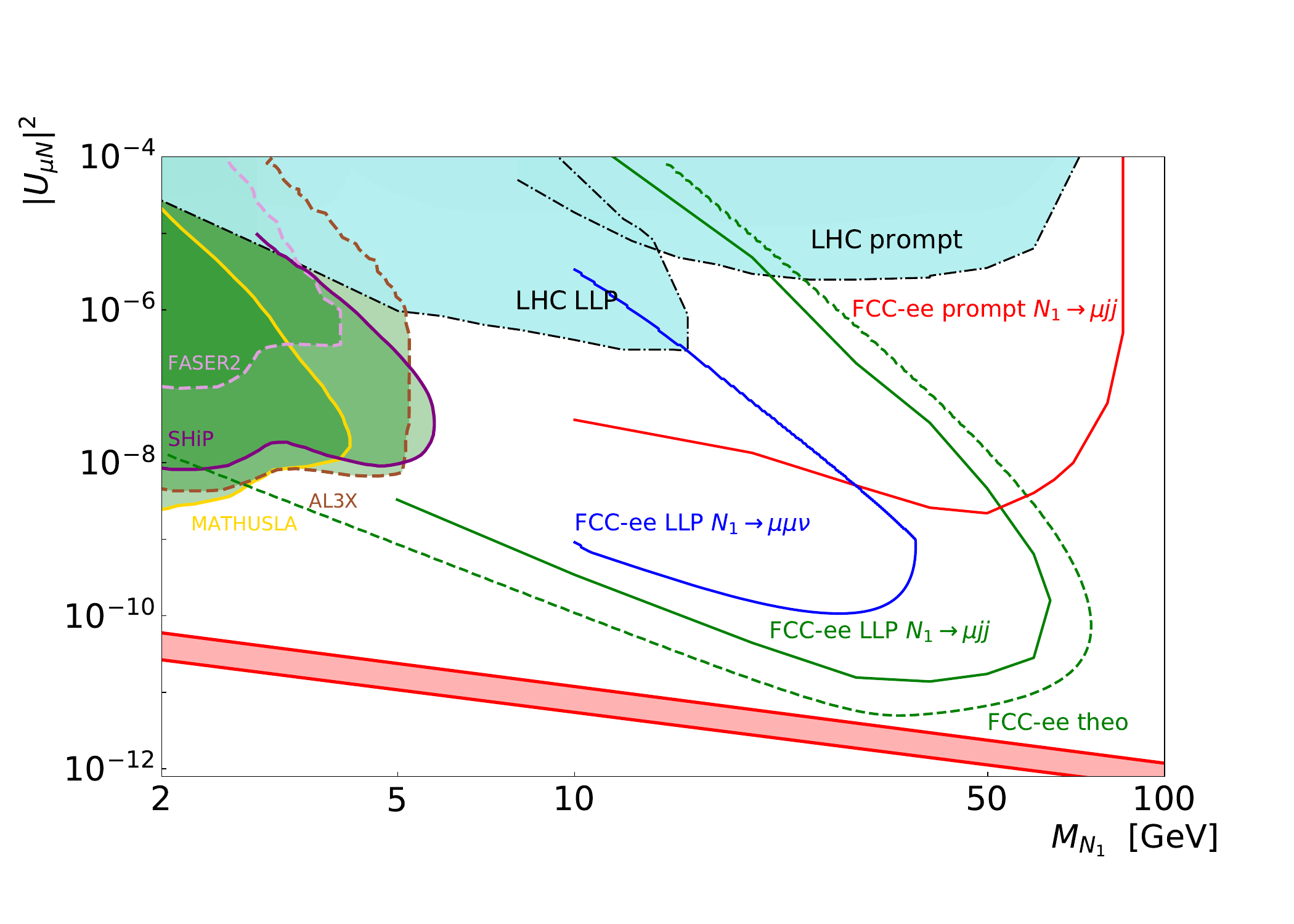} 
\caption{Discovery potential in the $\mhn-U_{\PGm}^2$ plane. The FCC-ee potential shown as a 
red (green) line for the prompt (long-lived) semileptonic analyses described  in the text. 
The blue line shows the reach of the leptonic decay $N_1\rightarrow\mu^+\mu^-\nu$. 
The dashed green line bounds the area where, out of $6\times10^{12}$ $Z$ bosons, three events
with visible HNL decays inside the full IDEA  detector are produced (based on the analytical 
formulas in \cite{Drewes:2022rsk}). 
The existing limits from LHC searches are given as turquoise areas. The expected discovery
potential of projected experimental searches based on long baseline experiments is shown as green
areas and is taken from the website accompanying \cite{Bolton:2019pcu}, where all the original work is cited.
} \label{fig::HNLMU2::sensitivityres}
\end{figure}

The results of the present study are compared with the existing experimental limits and with the expected limits for several proposed beam dump experiments in Figure \ref{fig::HNLMU2::sensitivityres}. 
The theoretical curve corresponding to three HNL decaying within the volume of a detector with 4.5~m diameter and 11~m length is also shown.
Thanks to the high Z statistics, these analyses could cover couplings down to $U_{\PGm}^2 \sim 10^{-11}$ for the kinematically accessible range of HNL masses.

%% file: hnlmu.bbl
\begin{thebibliography}{10}

\bibitem{FCC:2018evy}
A.~Abada et~al.
\newblock {FCC-ee: The Lepton Collider}: {Future Circular Collider Conceptual
  Design Report Volume 2}.
\newblock {\em Eur. Phys. J. ST}, 228(2):261--623, 2019.

\bibitem{Blondel:2014bra}
Alain Blondel, E.~Graverini, N.~Serra, and M.~Shaposhnikov.
\newblock {Search for Heavy Right Handed Neutrinos at the FCC-ee}.
\newblock {\em Nucl. Part. Phys. Proc.}, 273-275:1883--1890, 2016.

\bibitem{Blondel:2022qqo}
C.B. Verhaaren et~al.
\newblock {Searches for long-lived particles at the future FCC-ee}.
\newblock {\em Front. in Phys.}, 10:967881, 2022.

\bibitem{Abdullahi:2022jlv}
Asli~M. Abdullahi et~al.
\newblock {The present and future status of heavy neutral leptons}.
\newblock {\em J. Phys. G}, 50(2):020501, 2023.

\bibitem{nielsenth}
Sissel~Bay Nielsen.
\newblock Prospects of sterile neutrino search with the fcc-ee, 2017.
\newblock University of Copenaghen Master Thesis available at
  \url{https://nbi.ku.dk/english/theses/masters-theses/sissel-bay-nielsen/SisselBayNielsen_MastersThesis.pdf}.

\bibitem{Ding:2019tqq}
Jian-Nan Ding, Qin Qin, and Fu-Sheng Yu.
\newblock {Heavy neutrino searches at future $Z$-factories}.
\newblock {\em Eur. Phys. J. C}, 79(9):766, 2019.

\bibitem{Shen:2022ffi}
Yin-Fa Shen, Jian-Nan Ding, and Qin Qin.
\newblock {Monojet search for heavy neutrinos at future Z-factories}.
\newblock {\em Eur. Phys. J. C}, 82(5):398, 2022.

\bibitem{IDEAStudyGroup:2025gbt}
The IDEA~Study Group.
\newblock The idea detector concept for fcc-ee.
\newblock {\em arXiv:2502.21223}, 2025.

\bibitem{winter2023samples}
\url{https://fcc-physics-events.web.cern.ch/fcc-ee/delphes/winter2023/idea/}.

\bibitem{Atre:2009rg}
Anupama Atre, Tao Han, Silvia Pascoli, and Bin Zhang.
\newblock {The Search for Heavy Majorana Neutrinos}.
\newblock {\em JHEP}, 05:030, 2009.

\bibitem{Alva:2014gxa}
Daniel Alva, Tao Han, and Richard Ruiz.
\newblock {Heavy Majorana neutrinos from $W\gamma$ fusion at hadron colliders}.
\newblock {\em JHEP}, 02:072, 2015.

\bibitem{Degrande:2016aje}
Celine Degrande, Olivier Mattelaer, Richard Ruiz, and Jessica Turner.
\newblock {Fully-Automated Precision Predictions for Heavy Neutrino Production
  Mechanisms at Hadron Colliders}.
\newblock {\em Phys. Rev. D}, 94(5):053002, 2016.

\bibitem{Alwall:2014hca}
J.~Alwall, R.~Frederix, S.~Frixione, V.~Hirschi, F.~Maltoni, O.~Mattelaer,
  H.~S. Shao, T.~Stelzer, P.~Torrielli, and M.~Zaro.
\newblock {The automated computation of tree-level and next-to-leading order
  differential cross sections, and their matching to parton shower
  simulations}.
\newblock {\em JHEP}, 07:079, 2014.

\bibitem{Sjostrand:2014zea}
Torbj\"orn Sj\"ostrand, Stefan Ask, Jesper~R. Christiansen, Richard Corke,
  Nishita Desai, Philip Ilten, Stephen Mrenna, Stefan Prestel, Christine~O.
  Rasmussen, and Peter~Z. Skands.
\newblock {An introduction to PYTHIA 8.2}.
\newblock {\em Comput. Phys. Commun.}, 191:159--177, 2015.

\bibitem{deFavereau:2013fsa}
J.~de~Favereau, C.~Delaere, P.~Demin, A.~Giammanco, V.~Lema\^\i{}tre,
  A.~Mertens, and M.~Selvaggi.
\newblock {DELPHES 3, A modular framework for fast simulation of a generic
  collider experiment}.
\newblock {\em JHEP}, 02:057, 2014.

\bibitem{winter2023setup}
\url{https://github.com/HEP-FCC/FCC-config/tree/winter2023/FCCee}.

\bibitem{Bedeschi:2024uaf}
Franco Bedeschi.
\newblock {A vertex fitting package}.
\newblock {\em arXiv:2409.19326}, 9 2024.

\bibitem{Cacciari:2011ma}
Matteo Cacciari, Gavin~P. Salam, and Gregory Soyez.
\newblock {FastJet User Manual}.
\newblock {\em Eur. Phys. J.}, C72:1896, 2012.

\bibitem{ATLASstat}
ATLAS.
\newblock Formulae for estimating significance, 2020.
\newblock ATLAS Note ATL-PHYS-PUB-2020-025 available at
  \url{https://cds.cern.ch/record/2736148/files/ATL-PHYS-PUB-2020-025.pdf}.

\bibitem{Drewes:2022rsk}
Marco Drewes.
\newblock {Distinguishing Dirac and Majorana Heavy Neutrinos at Lepton
  Colliders}.
\newblock {\em PoS}, ICHEP2022:608, 2022.

\bibitem{Bolton:2019pcu}
Patrick~D. Bolton, Frank~F. Deppisch, and P.~S. Bhupal~Dev.
\newblock {Neutrinoless double beta decay versus other probes of heavy sterile
  neutrinos}.
\newblock {\em JHEP}, 03:170, 2020.

\end{thebibliography}
